# Pathology Steered Stratification Network for Subtype Identification in Alzheimer's Disease

**Running Title:** Stratification Network for Alzheimer's Disease

Enze Xu[1,#], Jingwen Zhang[1,#], Jiadi Li[2], Qianqian Song[3], Defu Yang[4], Guorong Wu[4,5], and Minghan Chen[1,*]

[1]Department of Computer Science, Wake Forest University, Winston-Salem, NC 27109, U.S.

[2]Department of Psychology, Wake Forest University, Winston-Salem, NC 27109, U.S.

[3]Department of Cancer Biology, Wake Forest School of Medicine, Winston-Salem, NC 27157, U.S.

[4]Department of Psychiatry, University of North Carolina at Chapel Hill, Chapel Hill, NC 27514, U.S.

[5]Department of Computer Science, University of North Carolina at Chapel Hill, Chapel Hill, NC 27599, U.S.

[#]Equal contribution

[*]Corresponding author: Minghan Chen. Email: chenm@wfu.edu. Full address: P.O. Box 7311, 1834 Wake Forest Road, Winston-Salem, NC 27109, U.S.

**Abstract**

**Background:** Alzheimer's disease (AD) is a heterogeneous, multifactorial neurodegenerative disorder characterized by three neurobiological factors beta-amyloid, pathologic tau, and neurodegeneration. There are no effective treatments for Alzheimer's disease at a late stage, urging for early detection and prevention. However, existing statistical inference approaches in neuroimaging studies of AD subtype identification do not take into account the pathological domain knowledge, which could lead to ill-posed results that are sometimes inconsistent with the essential neurological principles.

**Purpose:** Integrating systems biology modeling with machine learning, the study aims to assist clinical AD prognosis by providing a subpopulation classification in accordance with essential biological principles, neurological patterns, and cognitive symptoms.

**Methods:** We propose a novel pathology steered stratification network (PSSN) that incorporates established domain knowledge in AD pathology through a reaction-diffusion model, where we consider non-linear interactions between major biomarkers and diffusion along brain structural network. Trained on longitudinal multimodal neuroimaging data, the biological model predicts long-term evolution trajectories that capture individual characteristic progression pattern, filling in the gaps between sparse imaging data available. A deep predictive neural network is then built to exploit spatiotemporal dynamics, link neurological examinations with clinical profiles, and generate subtype assignment probability on an individual basis. We further identify an evolutionary disease graph to quantify subtype transition probabilities through extensive simulations.

**Results:** Our stratification achieves superior performance in both inter-cluster heterogeneity and intra-cluster homogeneity of various clinical scores. Applying our approach to enriched samples of aging populations, we identify six subtypes spanning AD spectrum, where each subtype exhibits a distinctive biomarker pattern that is consistent with its clinical outcome.

**Conclusions:** The proposed PSSN (1) reduces neuroimage data to low-dimensional feature vectors, (2) combines AT[N]-Net based on real pathological pathways, (3) predicts long-term biomarker trajectories, (4) stratifies subjects into fine-grained subtypes with distinct neurological underpinnings. PSSN provides insights into pre-symptomatic diagnosis and practical guidance on clinical treatments, which may be further generalized to other neurodegenerative diseases.

## 1. INTRODUCTION

Alzheimer's disease (AD), one of the common neurodegenerative disorders, causes progressive memory loss, cognitive decline, behavioral disturbance, functional dependence, and ultimately death[1]. No effective treatments for AD have been found so far, urging the need for early diagnosis and intervention. Current diagnostic systems for AD primarily rely upon clinical signs and symptoms. However, neuropsychological assessments alone are inadequate to reflect the underlying pathophysiological progressions considering the long preclinical period and diverse symptoms across individuals[2–4]. Additionally, post-mortem histological examination of AD pathology in brain tissue samples often does not align closely with clinical diagnosis[5]. This body of evidence further underscores the significant gap between symptomatic manifestations and latent development of AD pathology[6–8]. Therefore, to facilitate accurate prognosis and prompt intervention of AD, it is critical to stratify the aging population into fine-grained subtypes that are not only closely correlated with clinical outcomes but also characterize biomarker progression of cognitive decline.

According to the 2018 NIA-AA (National Institute on Aging and Alzheimer's Association) research framework[9], accumulations of beta-amyloid plaques (A) and pathologic tau (T), along with neuro-degeneration ([N]), are considered the three pathological hallmarks of Alzheimer's disease progression.

With recent advancement and accessibility of neuroimaging techniques, spatial data for biomarkers and atrophy patterns becomes increasingly available for clinical and research purposes. Striking efforts have been made to understand and explain the heterogeneity in AD progression from neuroimaging data[10–13], among which statistical inference, mostly machine learning approaches, have proven its promising potential in AD subtype identification[14–17]. For example, a Subtype and Stage Inference (SuStaIn) was developed to predict temporal patterns and identify AD subtypes from cross-sectional magnetic resonance imaging (MRI) progression patterns[18]. Vogel et al. extended the SuStaIn to include tau biomarker and identified four distinct progression subtypes[19]. Recently, a semi-supervised clustering via generative adversarial networks (Smile-GAN) was proposed to cluster subjects with distinct progression pathways using brain atrophy measurements[20]. However, those methods including most existing machine learning analyses do not take into account the underlying pathological mechanism of AD, which may result in ill-posed results that are inconsistent with essential biological laws.

Different from statistical inference approaches, mathematical modeling predicts disease progression based on existing knowledge of AD pathology[21]. Several models have been proposed to simulate the propagation pattern of biomarkers. For example, an ordinary differential equation (ODE) system was constructed to mathematically model the temporal progression of amyloid and neuronal dysfunction[22]. Following the diffusive nature of amyloid and tau, network diffusion models were developed to predict longitudinal patterns of atrophy and metabolism in AD across brain networks[23,24]. This pioneering work has shown the potential of capturing macroscopic properties of disease progression from a systems biology perspective. Hao et al. proposed a PDE-based (partial differential equation) system biology model to investigate the complex molecular pathways underlying AD[25]. But this model was not designed on a real brain domain and was restricted to theoretical analysis due to the lack of real data validation. Recently, a network-guided reaction-diffusion model was proposed to integrate both the interactions between AT[N] biomarkers and the diffusions along structural brain networks, which encodes the prevailing pathological mechanisms of AD and captures biomarker dynamics of individuals[26]. While those studies demonstrated the role of neurobiological factors in heterogeneous trajectories of cognitive decline[27–29], a comprehensive analysis of phenotypic heterogeneity is still lacking, which limits their applicability to the general population.

To disentangle the heterogenous neurodegeneration trajectories, we propose a novel pathology steered stratification network (PSSN) based on the combined power of data-driven deep learning and theory-based biological modeling. Our approach is integrated with existing neuropathological mechanisms and is evaluated on significant longitudinal multimodal imaging scans covering the full spectrum of cognitive states (from cognitive normal to AD), allowing the model to be trained on different progression paths. To the best of our knowledge, the proposed computational framework is the first subtype identification tool that jointly considers the biomarker pathological interactions, brain network diffusions, and clinical assessments for population stratification. Leveraging theory-based biological models and data-driven deep learning, our PSSN: incorporates neuropathological domain knowledge to ensure neurologically consistent results; ameliorates the limitation of sparse longitudinal imaging data by learning the spatiotemporal dynamics of AT[N] biomarkers; stratifies subjects into fine-grained subtypes with distinct neurological underpinnings and phenotypic outcomes; characterizes subtype transition path on AD spectrum, providing insight into pre-symptomatic prognosis and prevention. It is important to note that fully validating the subtypes identified by our model may be challenging. This is primarily due to the absence of a definitive ground truth for subtype assignment in Alzheimer's disease and the inherent heterogeneity present in the AD data. Nevertheless, we believe that our model could provide insights and contributes to the ongoing efforts to understand the complex nature of AD progression. We will continue to process additional temporal neuroimaging data to enhance the representativity of our model and improve its accuracy.

## 2 MATERIALS AND METHODS

### 2.1 Data Collection and Image Processing

The data used in this study was collected from Alzheimer's Disease Neuroimaging Initiative (ADNI) database. We retrieved (i) demographic information, including age, gender, education, and occupation information; (ii) clinical assessment from a wide range of neuropsychological domains, including memory, executive, language, sociability, attention, etc.; (iii) longitudinal neuroimaging data, including Amyloid-, Tau-, and FDG-PET for AT[N] biomarker burden, and DWI for structural brain networks. Subjects selected here have two or more longitudinal scans and clinical assessment data. 320 subjects were then selected based on three criteria: (i) have available Amyloid-PET, Tau-PET, FDG-PET, T1-weighted MRI, and DWI scans, (ii) have at least one follow-up PET scan of A, T, or [N] biomarkers to track AD progression, and (iii) have a clinical diagnostic label indicating their cognitive status, either as cognitive normal (CN) or Alzheimer's disease (AD), assigned for each PET scan.

2.1.1 Brain Network Construction

DWI images were aligned with each subject's T1-weighted MRI. We first parcellate brains to 148 regions based on the parcellation framework proposed in [40] and then applied the surface seed-based probabilistic fiber tractography in FreeSurfer[41]. The number of fibers connecting two brain regions is counted to measure the strength of the anatomical connectivity of pair-wise regions and stored in the connectivity matrix. A total of 506 structural brain networks were constructed from DWI images.

### 2.1.2 Regional AT[N] Level Acquisition

2,807 amyloid PET (1,252 subjects), 1,009 tau PET (670 subjects), and 3,590 FDG PET (1,516 subjects) images were parcellated into 148 regions using Destrieux atlas[40] by aligning with each subject's T1-weighted MR image. We then calculate the regional standard uptake value ratio (SUVR) for amyloid-, tau-, and FDG-PET to represent the regional value of corresponding biomarkers, which is normalized by the whole cerebellum reference. All the PET data we collected have passed the quality control of the ADNI.

### 2.1.3 Cognitive Reserve

A computational proxy of the cognitive reserve[36] is used here to quantify the subject-specific network resilience to AT[N] pathology. Subjects' demographic data, socioeconomic factors, cerebrospinal fluid biomarkers, and AD-related polygenetic risk are individually evaluated and serve as a moderator against neurodegeneration in our model.

### 2.1.4 Clinical Assessments

MMSE, ADAS, and CDR scores are indicators of subjects' overall cognition status and are used to guide our PSSN. We use Everyday Cognition Questionnaire (ECog) scores to validate and analyze our results. ECog is scored on a four-point scale where higher scores represent more severe dementia, based on one global factor and six domain-specific factors: everyday memory (ECogMem), language (ECogLang), visuospatial abilities (ECogVisspat), planning (ECogPlan), organization (ECogOrgan), and divided attention (ECogDivatt). All the above clinical assessment data were collected from the ADNI database.

## 2.2 Reaction-Diffusion Model

Based on existing neuropathological knowledge, we first construct a reaction-diffusion model of Alzheimer's disease to characterize the spatiotemporal interaction and diffusion of AT[N] biomarkers. This biological model can predict subject-specific long-term propagation patterns across brain networks, which enables PSSN to identify AD subtypes following neurological principles.

### 2.2.1 Model Entity

As shown in Fig. 1(a), the model takes four entities of AD as input: (i) A biomarker level measured from amyloid PET at 148 brain regions, which is a vector of 148×1; (ii) T biomarker level measured from tau PET of size 148×1; (iii) [N] biomarker level measured from FDG PET of size 148×1; Each AT[N] input is a 592×M matrix, where M denotes the number of subjects included in our model. (iv) brain network (graph Laplacian), which is a 148×148 matrix used as the diffusive pathway for AT[N]. These multimodal neuroimaging data jointly provide a comprehensive AD profile of each subject.

### 2.2.2 Model Design

In Fig. 1(b), the core of our reaction-diffusion model is the amyloid cascade hypothesis[30–34]. (i) Amyloid activates hyperphosphorylation of tau, tau triggers subsequent neurodegeneration, and damaged neurons release more amyloid to the brain (dashed arrows)[35]. (ii) amyloid and tau have constant production (solid arrows) and density-based degradation (hollow arrows). (iii) neuronal resilience moderates the rate of neurodegeneration[36]. (vi) amyloid and tau diffuse along the structural brain network via Laplacian operator[24,37,38]. This model quantitatively characterizes the spatiotemporal dynamics of AT[N] biomarkers (Fig. 1(c)), laying a solid foundation for AD prognostic subtype identification. Detailed descriptions can be found in [26].

## 2.3 Pathology Steered Stratification Network

Following our biological model, we construct a deep predictive neural network to disentangle the heterogenous neurodegeneration progression and stratify subjects into fine-grained subtypes within distinct neurobiological underpinnings.

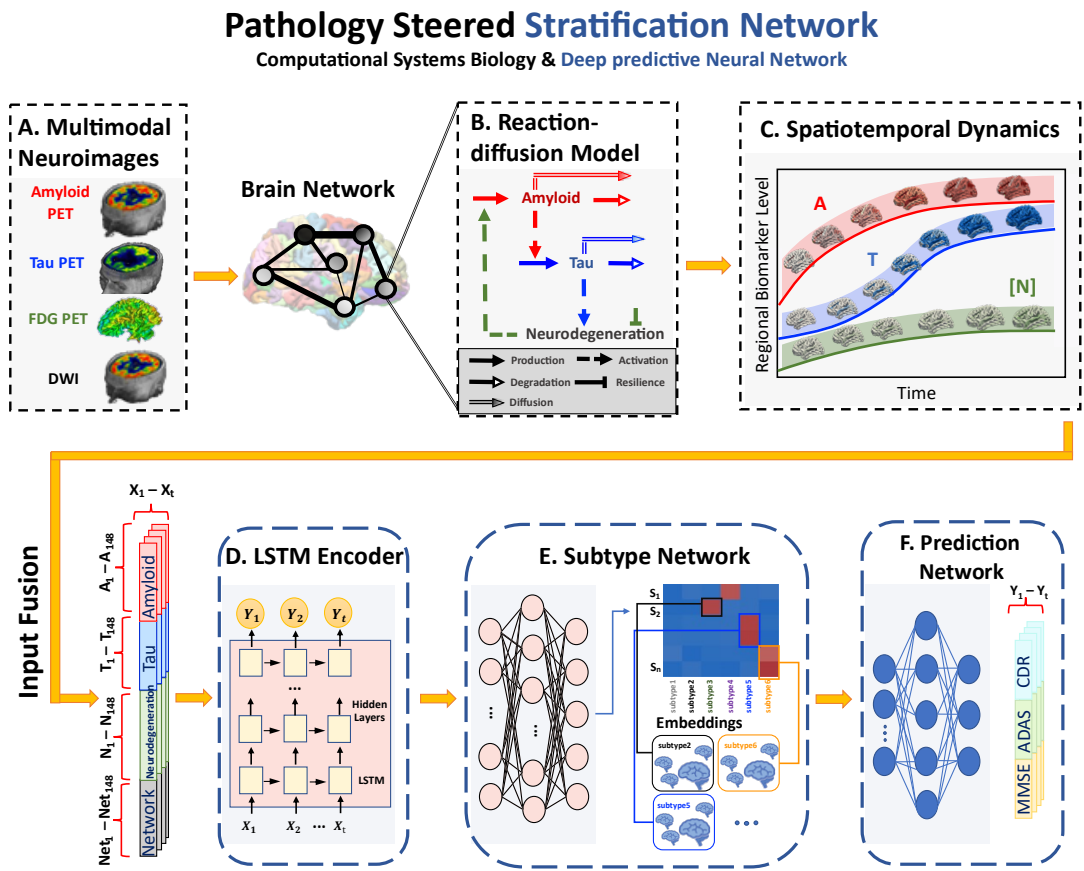

**Fig. 1. Pathology Steered Stratification Framework: A novel integration of a systems biology model and deep predictive neural network. (a)** Longitudinal multimodal neuroimages input: Amyloid-, Tau-, FDG-PET, and DWI scans are processed to indicate regional amyloid, tau, neurodegeneration, and network connectivity level (AT[N]-Net), respectively. **(b)** Reaction-diffusion Model: a diffusive AT[N] cascade model was proposed based on several canonical AD pathways. **(c)** Spatiotemporal Dynamics: reaction-diffusion model generates subject-specific long-term AT[N] trajectory prediction. **(d)** Feature Extraction Network: a LSTM network learns a lower-dimension representative feature from the predicted trajectories and structural network. **(e)** Subtype Network: a fully connected neural network that classifies the learned features to the subtype-probability vectors. **(f)** Prediction Network: a fully connected neural network that predicts clinical assessments using the subtype-probability vectors.

2.3.1 Data Fusion

Let $t_m$ denote the number of clinical assessments for subject $m$. For each subject $m$, we (1) predict the trajectories of AT[N] biomarkers using the reaction-diffusion model; (2) select $t_m$ points from subject $m$'s trajectory corresponding to the time points of clinical assessments, concatenate AT[N] data $\{A_1, \ldots, A_{148}, T_1, \ldots, T_{148}, N_1, \ldots, N_{148}\}$ and the diagonal elements of Laplacian matrix $\{L_1, \ldots, L_{148}\}$ for each $t \in \{1, \ldots, t_m\}$ to generate a $592 \times t_m$ matrix, which serves as the input to our stratification network; and (3) extract clinical assessments of subject $m$ as output, which includes mini-mental state examination (MMSE), Alzheimer's disease assessment scale (ADAS), and clinical dementia rating (CDR). Thus for a total population of $M$, we have an input matrix $X = \{X_m^t | m = 1, \ldots, M, t = 1, \ldots, t_m\}$, where each $X_m^t$ represents the AT[N]-Net data of subject $m$ at visit $t$, and output matrix $Y = \{Y_m^t | m = 1, \ldots, M, t = 1, \ldots, t_m\}$, where $Y_m^t$ represents the three clinical scores of subject $m$ at visit $t$.

2.3.2 Stratification Network

We aim to stratify aging brains into a set of subtypes, written as $S = \{S_m^t | m = 1, \ldots, M, t = 1, \ldots, t_m\}$, where $S_m^t$ is the subtype index for subject $m$ at the $t^{th}$ visit. To distinct neurobiological underpinnings between subtypes as well as preserve the shared symptoms and cognitive decline patterns within each subtype, three sub-networks are proposed as follows.

1. Feature Extraction Network (Fig. 1(d)). The feature extraction network is a many-to-many Long Short-Term Memory (LSTM) network to learn representative features of each sample. For each input sample $X_m^t$ of subject $m$ at the $t^{th}$ visit, the corresponding AT[N]-Net data is reduced to a feature vector in a lower-dimensional space, represented as an AT[N]-Net feature vector $P_m^t = P_\theta(X_m^1, X_m^2, \ldots, X_m^t)$, $P_\theta$ is the feature representation learning network, and $\theta$ denotes the parameters of the network.

2. Subtype Network (Fig. 1(e)). The subtype network is a fully connected neural network that learns patterns in the subject pool and then maps each AT[N]-Net feature vector $P_m^t$ into a subtype-specific feature vector $C_m^t(k)$, where $k \in \{1,2,\ldots,K\}$ and $K$ denotes the number of identified subtypes. Also, we include fuzzy assignment by minimizing the entropy-based regularization term $E_m^t = -\sum_{k=1}^{K} C_m^t(k) log(C_m^t(k))$ to avoid the trivial stratification solution (one dominant cluster). Let $\psi$ denote the parameters of subtype network.

3. Prediction Network (Fig. 1(f)). We use a fully connected neural network to predict each subject's MMSE, CDR, and ADAS scores based on the subtype assignment probability $C_m^t$ generated in Subtype Network. A $L_2$-norm loss function is used to measure the prediction error, written as $l_m^t = \left\| Y_m^t - \hat{Y}_m^t \right\|^2$. This clinical assessment-guided feature allows us to group subjects with similar clinical stages. Let $\phi$ denote the parameters of the prediction network.

### 2.2.3 Optimization

Three sub-networks are jointly updated by the loss function

$$L(\theta, \psi, \phi) = \sum_{m=1}^{M} \sum_{t=1}^{t_m} \left( l_m^t + \lambda \cdot E_m^t(k) \right),$$

where $l_m^t$ allows clinical assessments to guide the stratification process and $\lambda$ is a hyperparameter used to control the level of fuzzy assignment. Directly calculating the backpropagation derivative is quite complicated in the Stratification network, as it aims to minimize the divergence between clinical scores and predictions as well as to find the optimal cluster assignments at the same time. Thus, we apply the actor-critic model [39] for optimization, which iteratively optimizes two groups of sub-networks.

Let $\nabla_\phi, \nabla_\theta, \nabla_\psi$ represent the gradients of the backpropagation process, respectively. We first fix the parameter of the prediction network ($\phi$) to estimate parameters of feature extraction and subtype network ($\theta, \psi$). By fixing $\phi$, gradients $\nabla_\theta$ can be given by:

$$\nabla_\theta = \frac{\partial L}{\partial \theta} = \sum_{m=1}^{M} \sum_{t=1}^{t_m} \left( \sum_{k=1}^{K} \left\| Y_m^t - \hat{Y}_m^t \right\|^2 \nabla_\theta E_m^t(k) - \lambda \nabla_\theta \sum_{k=1}^{K} E_m^t(k) log(E_m^t(k)) \right)$$

Similarly, the gradients $\nabla_\psi$ can be given by:

$$\nabla_\psi = \frac{\partial L}{\partial \psi} = \sum_{m=1}^{M} \sum_{t=1}^{t_m} \left( \sum_{k=1}^{K} \left\| Y_m^t - \hat{Y}_m^t \right\|^2 \nabla_\psi E_m^t(k) - \lambda \nabla_\psi \sum_{k=1}^{K} E_m^t(k) log(E_m^t(k)) \right)$$

At each stage, we update the subtype assignment by the updated subtype-specific features $\{C_m^t\}$. Then by fixing the feature extraction network and subtype network $(\theta, \psi)$, we iteratively update the prediction network whose gradient is given by:

$$\nabla_\phi = \frac{\partial L}{\partial \phi} = \sum_{m=1}^{M} \sum_{t=1}^{t_m} \frac{\partial l_m^t}{\partial \phi}$$

## 3 RESULTS

Our PSSN outputs two types of information. First, the PSSN uses a reaction-diffusion model to predict the spatiotemporal dynamics of AT[N] biomarkers for each subject. This model incorporates current neuropathological pathways between these biomarkers and integrates their diffusion across brain networks. By doing so, we ensure that the prediction results are aligned with known neuropathological pathways. The predicted spatiotemporal dynamics provide insights into how the AT[N] biomarkers evolve and interact over time within the aging brain. This information helps us understand the underlying biological processes and mechanisms associated with the progression of neurodegenerative diseases. Secondly, utilizing the learned AT[N] evolution trajectories, the deep predictive neural network of PSSN is able to stratify aging brains into fine-grained subtypes. These subtypes are identified based on the distinctive patterns of biomarker dynamics observed across different individuals. By capturing the heterogeneity within the aging population, this stratification enhances our understanding of the diverse pathways and trajectories of neurodegeneration. It is particularly valuable when dealing with sparse longitudinal imaging data, addressing the limitations of data availability. PSSN allows us to characterize and study subgroups of individuals with similar biomarker profiles and clinical outcomes, providing opportunities for targeted interventions and personalized treatment approaches.

### 3.1 Experiment Setup and Parameter Tuning

We conducted our experiments on a multi-core Redhat Linux machine with two 32GB NVIDIA Tesla V100 GPUs. Our model was developed and constructed using TensorFlow, a popular machine learning framework. To train our PSSN, the data ($X$ and $Y$) is randomly split into training (80%) and validation (20%) sets. Since there is no ground truth for subtype assignments, we determine the optimal number of clusters by jointly considering cluster results from K-means, SuStaIn, and our PSSN. Taking K-means and SuStaIn as baselines, our method attains the lowest cluster variance ratio with respect to both K-means SuStaIn methods. Indeed, when $K = 6$, PSSN yields the lowest variance for all ECog scores except ECogOrgan ($K = 5$), as shown in Fig. 2.

The hyper-parameters were set to $K = 6$, $\lambda = 10^{-5}$, $FC\_dim = 8$ and $keep\_prob = 0.7$, where K represents the number of subtypes and has the most significant impact on both the clustering outcomes and final experimental results; $\lambda$ is the parameter for calculating the loss function; $FC\_dim$ is the number of layers of the fully connected networks used in both the Subtype Network and the Prediction Network; $keep\_prob$ is the probability of data kept at each dropout.

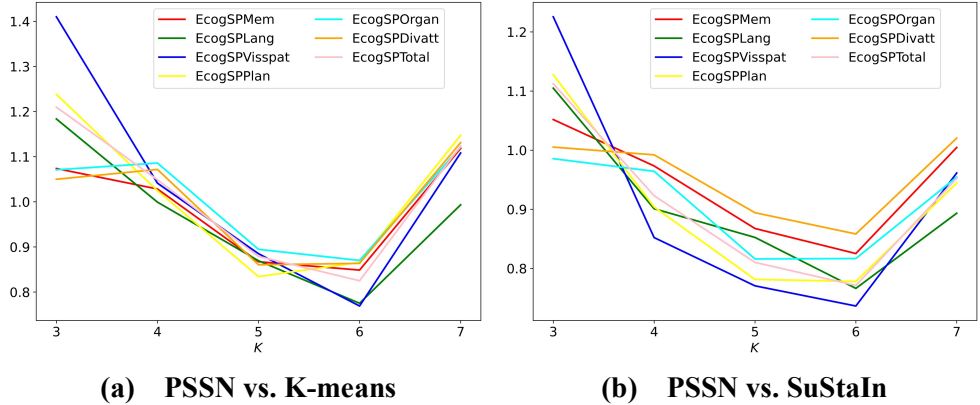

(a) PSSN vs. K-means      (b) PSSN vs. SuStaIn

**Fig. 2. Optimal number of subtypes.** Intra-cluster variance ratios on ECog scores between **(a)** PSSN and K-means, **(b)** PSSN and SuStaIn with varying $K$, respectively. For instance, as shown in subplot (a), the EcogSPVisspat (blue line) has a value of 0.8850 at $K = 5$. This indicates that utilizing the PSSN can lead to an 11.50% decrease in intra-cluster variance when compared to directly applying K-means.

## 3.2 Stratification Comparison

We use seven ECog scores (Mem, Lang, Visspat, Plan, Organ, Divatt, and Total) to evaluate the subtype stratification results as they reflect the clinical stages of AD progression and are unknown to all models. We first analyze the distribution of subtype classification in three methods. As shown in Table 1, K-means and SuStaIn identified a giant cluster that includes more than 50% of observations despite extensive attempts. One possible explanation is their incapability to handle high-dimensional spatial data, where our input dimension is 148 brain regions. The elastic feature extraction network in PSSN can adapt to time series data with various lengths. Given the sporadic availability of neuroimaging data in the real world, such a design maximizes the utility of available data.

### 3.2.1 Intra-cluster Homogeneity

We then investigate the consistency of clinical outcomes within each cluster. For each subtype identified by the PSSN, we calculate the variance of clinical scores to evaluate the quality of subtype classification, where a small variance indicates intra-cluster homogeneity. Fig. 3(a) shows the intra-cluster variance of seven ECog scores by K-means (top), SuStaIn (middle), and PSSN (bottom). Note that each column of the heap map is individually normalized using column-wise z-scores, ensuring that each method is compared using the same mean and standard deviation values. Our PSSN method achieves superior intra-cluster consistency compared to K-means and SuStaIn in all ECog assessments.

**Table 1.** Cluster sizes of K-means, SuStaIn, and PSSN methods.

|  | K-means | SuStaIn | PSSN |
|---|---|---|---|
| **Subtype #1** | 275 | 207 | 211 |
| **Subtype #2** | 1115 | 238 | 217 |
| **Subtype #3** | 76 | 304 | 423 |
| **Subtype #4** | 237 | 861 | 743 |
| **Subtype #5** | 246 | 327 | 210 |
| **Subtype #6** | 10 | 22 | 155 |

3.2.2 Inter-cluster Heterogeneity

A pairwise $t$-test was performed to check the inter-cluster variance. In Fig. 3(b), we plot a triangle with $C_6^2 = 15$ grids where each grid represents the results of two-sample $t$-tests. As shown in Fig. 3(b), our PSSN method achieves the best performance since it has the least biased subtypes (pink grids, < 5% of the subject pool) and the greatest number of pairs with significant inter-cluster differences (red grids, $p$-value < 0.05). We further rank each subtype according to the severity of averaged MMSE, CDR, and ADAS scores. It is found that our PSSN has the smallest cross-ranking links, indicating the best consistency within each subtype (Fig. 3(c)). With the lowest intra-cluster variance as well as the highest inter-cluster difference, our PSSN achieves the best coherence of all clinical assessments and sets a solid cornerstone for its clinical implications and applications.

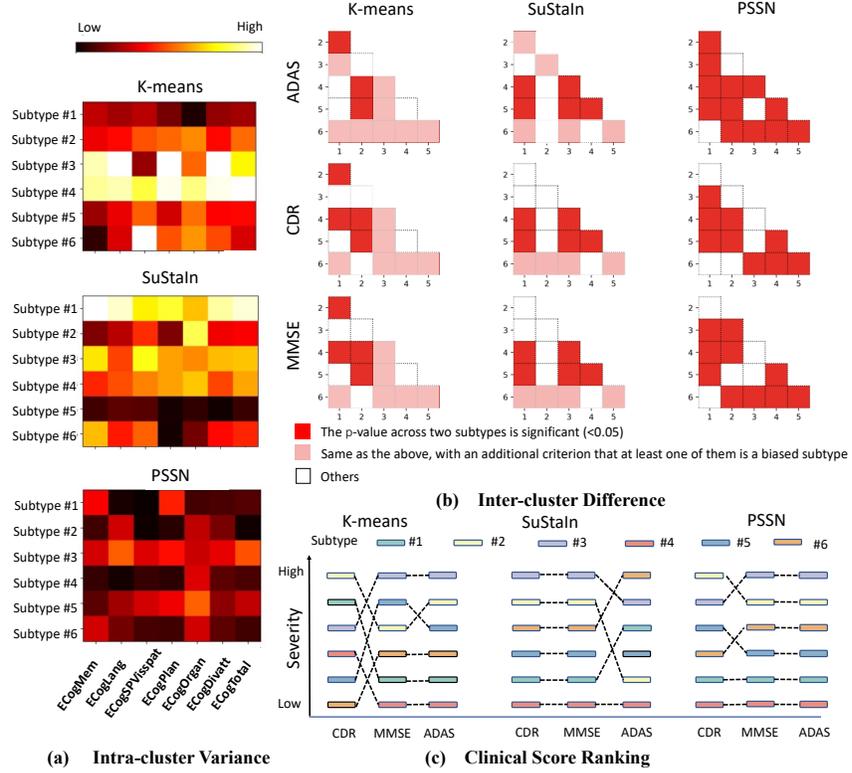

**Fig. 3. Intra-cluster homogeneity and inter-cluster heterogeneity. (a)** Intra-cluster Variance heatmap of ECog scores for PSSN, K-means and SuStaIn methods, where darker color represents lower variance. **(b)** Inter-cluster Variance: subtype pairs are marked red if the adjusted $p$-value across their MMSE, CDR, and ADAS scores is significant (< 0.05). For significant subtype pairs, the gird is marked pink if at least one of these two subtypes is a biased subtype. **(c)** Trajectory plots of subtypes ranking according to MMSE, CDR, and ADAS scores.

### 3.3 Clinical Insights

Current diagnoses of AD are often delayed due to the wide spectrum of clinical symptoms among individuals. Given the fine-grained subtype identified by our model, we further examine the distribution of ECog scores to explore the representative symptoms of each subtype.

3.3.1 ECog Entropy

To evaluate the inter-cluster difference of ECog scores, we calculate the cluster-wise entropy $E(L)$ in seven ECog scores:

$$E(L) = \{-\sum_{i=1}^{K} p(l_{i,j}) \log_2 p(l_{i,j}) \mid j = 1, \ldots, 7\},$$

where $j$ refers to the index of ECog scores, and $p(l_{i,j})$ is the probability of assigning the $j^{th}$ ECog scores to $i^{th}$ subtype. The average entropies of each ECog score are shown in Fig. 4(a). Compared to K-means and SuStaIn, our PSSN achieves the lowest entropies in all seven categories of ECog scores, showing the greatest stability in subtype identification.

| Method/ECog | Memory | Language | Visuospatial Abilities | Planning | Organization | Divided Attention | Total |
|---|---|---|---|---|---|---|---|
| PSSN | 0.4798 | 0.4616 | 0.4390 | 0.4313 | 0.4474 | 0.4422 | 0.5424 |
| K-means | 0.6255 | 0.6676 | 0.6052 | 0.5680 | 0.6182 | 0.5850 | 0.7074 |
| SuStaIn | 0.8084 | 0.7590 | 0.6463 | 0.6479 | 0.6718 | 0.6396 | 0.9041 |

(a) Entropy Comparison

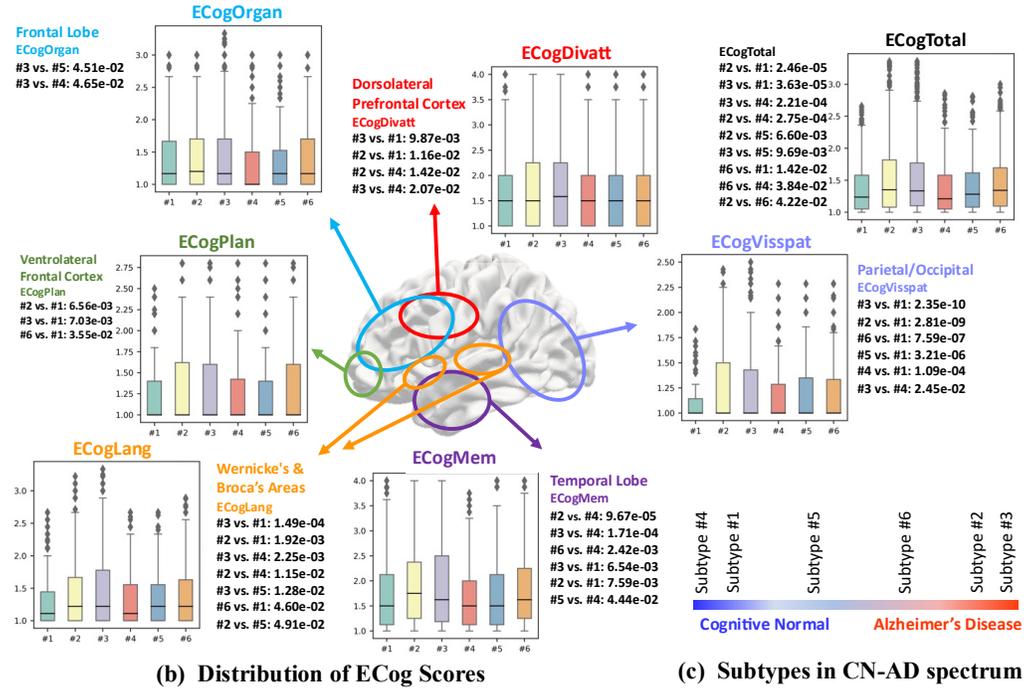

(b) Distribution of ECog Scores    (c) Subtypes in CN-AD spectrum

**Fig. 4. Clinical interpretation of stratification results.** (a) Entropy Comparison: ECog entropies across subtypes for K-means, SuStaIn, and PSSN methods. (b) Distribution of ECog scores. Each ECog score is mapped into a specific domain based on the functional parcellation of the human brain: temporal lobes regulate everyday memory (purple)[42], Wernicke's & Broca's areas regulate language (orange)[43], parietal and occipital lobes regulate visuospatial abilities (deep periwinkle)[44], ventrolateral frontal cortex regulates planning (green)[45], frontal lobe regulates organization (blue)[46], and dorsolateral prefrontal cortex regulates divided attention (red)[47]. Each boxplot shows the distribution of corresponding clinical assessments. The pairwise $t$-test is used to measure the difference between subtypes, and all pairs with < 0.05 adjusted $p$-values are displayed next to the boxplot. Note the subtype with a higher mean ECog score is listed on the left-hand side. (c) Relative stages of six subtypes on AD spectrum based on ECog scores.

### 3.3.2 Subtype Symptom

In Fig. 4(b), the six domains of ECog (except ECog-Total) are mapped to six brain domains based on functional parcellation (see Fig. 4 for detailed brain region parcellation). Further in Fig. 4(c), we conclude the position of subtypes in the continuous spectrum: CN-like subtypes (#1, #4), MCI-like subtypes (#5, #6), and AD-like subtypes (#2, #3). The observed pattern summarized below offers new insight into the pre-diagnosis and treatment of AD as it inherently combines physiological and psychological markers.

*AD State.* As we can see clearly from Fig. 4(b), subtype #3 and subtype #2 present distinct cognitive levels in almost all domains. Subtype #3 has significantly higher scores than (i) subtype #4 in six domains (Organ, Divatt, Visspat, Mem, Lang, and Total), (ii) subtype #1 in domains (Plan, Divatt, Visspat, Mem, Lang, and Total). Subtype #2 is significantly higher than (i) subtype #4 in Divatt, Mem, Lang, Total, (ii) subtype #1 in Divatt, Mem, Lang, Total, Plan, Visspat. This pattern also aligns with the predicted AT[N] level generated by our spatiotemporal model (discussed in Section *Neurobiological Evaluation*). Based on the consistency of pathological biomarkers and clinical scores, we consider subtypes #3 and #2 as AD types, where high-level AT[N] accumulates across the brain and multiple symptoms manifest together simultaneously. Although it is hard to differentiate subtypes #3 and #2 based on reported symptoms, we could observe distinct patterns of pathological burdens Fig. 5(a).

*CN State.* Subtypes #1 and #4 are on the other end of the AD spectrum as they have the lowest grades in almost all domains. A closer look at the clinical scores reveals that subtype #1 has particularly low scores on planning and visuospatial abilities, implying low AT[N] accumulations in the ventrolateral frontal cortex, parietal lobe, and occipital lobe, which is further validated by the subtypes' phenotypes shown in Fig. 5(a). Subtype #4 scores are significantly lower in memory, which is also consistent with the brain mapping of AT[N] biomarkers. A further examination shows that subtype #4 has more problems with visual-spatial abilities, which suggests subtype-specific treatments in parietal and occipital lobes and could be used in clinical settings to differentiate subtypes #4 and #1.

*MCI State.* Subtypes #5 and #6 are intermediate states. Subtype #5 scores significantly higher on memory than subtype #4 and on visual-spatial ability than subtype #1. Subtype #6 has higher scores in almost all cognitive domains than subtypes #1, #4, which implies subtype #6 should be placed closer to the AD end on cognitive continuums. Subtype #5 has close to normal organization and language abilities, while subjects in subtype #6 tend to score higher in planning, language abilities, and total assessments. Such distinction suggests heavier pathologic burdens in Subtype #6's Wernicke's and Broca's areas, which is confirmed in Fig. 5(a).

## 4 DISCUSSIONS

Due to the massive heterogeneity between neurobiological examination and clinical diagnosis, we are interested in the associated neurobiological factors for each identified subtype. We compute the subtype transition matrix, analyze the corresponding AT[N] pattern, and generalize an AD evolutionary graph for subtype prognosis.

*Transition Matrix.* As defined above, each subject is labeled as $s_m^t$ which is the subtype index (from 1 to $K = 6$) of subject $m$ at visit $t$. Since subjects may transit from one subtype to another, we record the occurrence of such transitions by ordered pairs defined as $Q = \{q_i | i = 1, \ldots, K\} = \{(s_m^t, s_m^{t+1}) | m = 1, \ldots, M, t = 1, \ldots, t_m - 1\}$. Each $q_i$ represents all the transition pairs starting in subtype #$i$. We then count the frequency of ordered pairs and calculate the row stochastic transition matrix $Pr = [pr_{i,j}], i, j = 1, \ldots, K$, where $pr_{i,j} = \frac{|\{(s_m^t, s_m^{t+1}) | s_m^t = i, \ s_m^{t+1} = j\}|}{|q_i|}$ is the possibility of subtype #$i$ transiting to subtype #$j$.

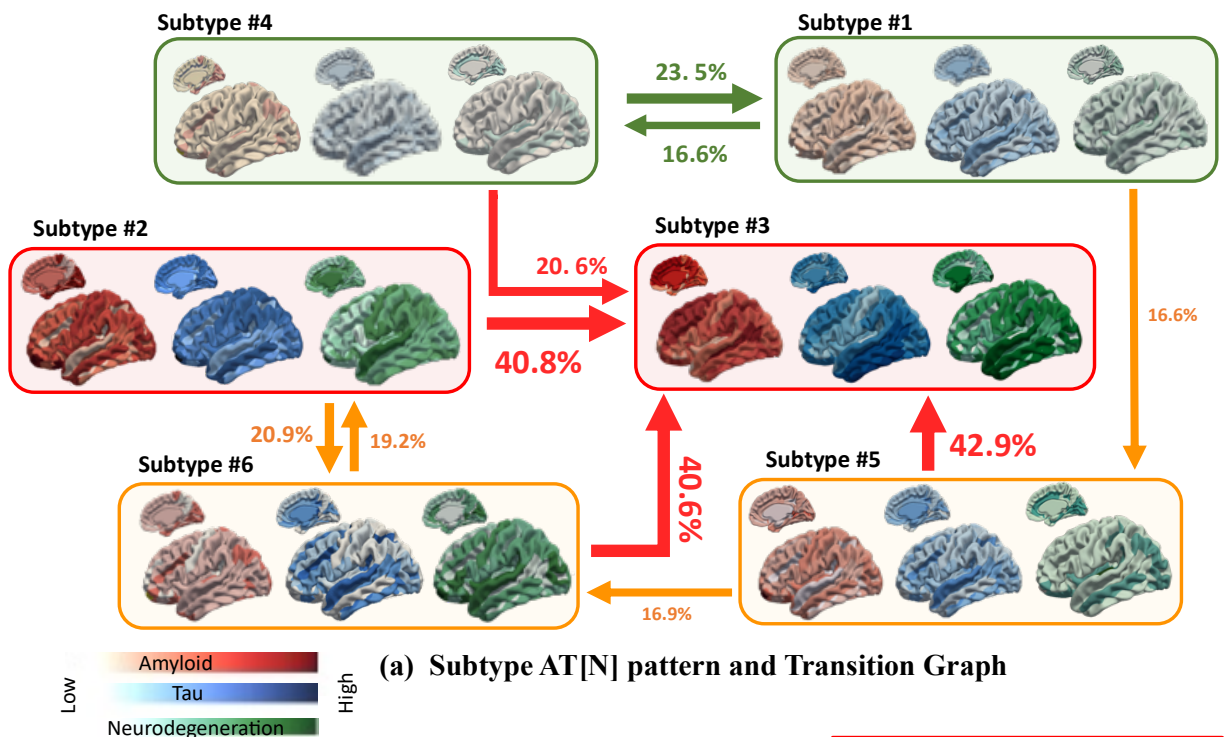

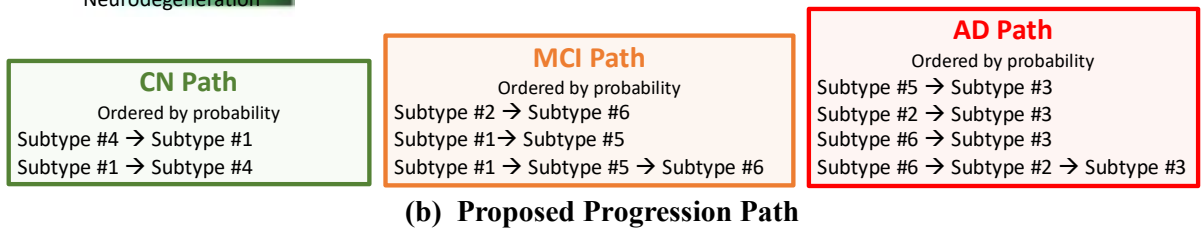

**Fig. 5. Hypothetical subtype transition graph in AD progression. (a)** The subtype conversion diagram was drawn upon the count of subjects converting from one subtype to another. The probability of transition is labeled next to each arrow and indicated by the arrow width. The inside rounded box shows the AT[N]

burden of each subtype, where the lightness of color represents the level of AT[N] biomarker (red for A, blue for T, green for [N]). Only transitions with probabilities > 16% were plotted in the diagram. **(b)** Subtype brains with low, intermediate, and high AT[N] values are mapped to CN-, MCI-, and AD-like states. We summarize possible CN/MCI/AD pathways based on starting and ending states.

*AT[N] Pattern.* We check the topological patterns of average AT[N] burden across brain networks to investigate the relationship with their clinical outcomes, indicated by AT[N] pattern (Fig. 5(a)) and subtype-typical clinical profiles (Fig. 4(b)). Consistent with normal cognitive functions reflected by ECog scores, AT[N] burdens are the lowest for subtype #4 except for parietal and occipital lobes, which matches the jump in visuospatial scores. For subtype #1, only mild AT[N] burden is observed across the brain. Despite relatively higher tau levels in the superior and inferior temporal gyrus, the neuron loss is not noticeable in the temporal lobe, which could explain subtype #1's significantly better memory abilities. Close attention should be paid to this cohort if they experience a sudden decline in memory. Subtype #5 resides in the intermediate state of AD spectrum. Subtype #6 is also in the intermediate state but closer to AD state with elevated tau and neurodegeneration level in temporal lobes, which accords with ECog memory score. Subtypes #2 and #3 are identified as AD state since they have significantly higher whole-brain AT[N] levels and clinical scores in all domains. Subtype #3 tends to harbor more neurodegeneration in Wernicke's area than subtype #2, including angular gyrus and supramarginal gyrus that are in the parietal lobe[48]. Thus, subtype #3 is more susceptible to declined comprehension ability, leading to severe survival problems in real-life situations[49].

In Fig. 5(a), we connect each subtype-wise average AT[N] pattern using the obtained transition matrix, where thicker arrows represent high transition probability. Subtype #4 has the lowest AT[N] level in the prefrontal cortex and has a 23.5% probability of converting to subtype #1, which accumulates more pathological burdens in the prefrontal cortex. Subtype #1 has a 16% probability of converting to the intermediate state, subtype #5, and then to subtype #6. These MCI states each has around 40% probability of developing into AD state, which is marked by further increments of AT[N] accumulations across the exocortex. Fig. 5(b) summarizes hypothetical CN, MCI, AD pathways, which are congruent with the order of subtypes on the continuum in Fig. 4(c), demonstrating the consistency between physiological and psychological signs using our pathology steered method. Clinicians should closely monitor their cognitive abilities and provide prompt intervention if there is a sudden deterioration.

*Final Stable State.* For the subtype transition matrix $Pr$, each subtype can be considered as a node in a graph, and the transition probability indicates the directed link between two subtypes. As the initial distribution of subtypes in our subject pool is $\pi^{(0)} = \{0.11,\ 0.11,\ 0.22, 0.38, 0.11, 0.08\}$ and $\pi^{(n)} = \pi^{(0)} Pr^n$, we can get the steady state vector $v$ of the transition matrix by taking the infinite limit of $n$, written as $v = \pi^{(0)} \cdot \lim_{n \to \infty} Pr^n$. Fig. 6 shows the transition matrix at different steps ($n = 1, 3, 5$) using heatmaps, where darker colors represent higher transition probabilities. When $n = 0$, the heatmap is the initial transition matrix. As we increase the exponent, the matrix approaches the stable state vector $v$. We can calculate $v$ by solving $Pr \cdot v = v$, which is the eigenvector of the transition matrix corresponding to the eigenvalue $\lambda = 1$. More than 50% of subjects will progress to AD states (subtypes #2 and #3), 33% of subjects will stay in MCI state (subtypes #5 and #6), and only 13% of subjects will stay in cognitive normal states (subtypes #1 and #4).

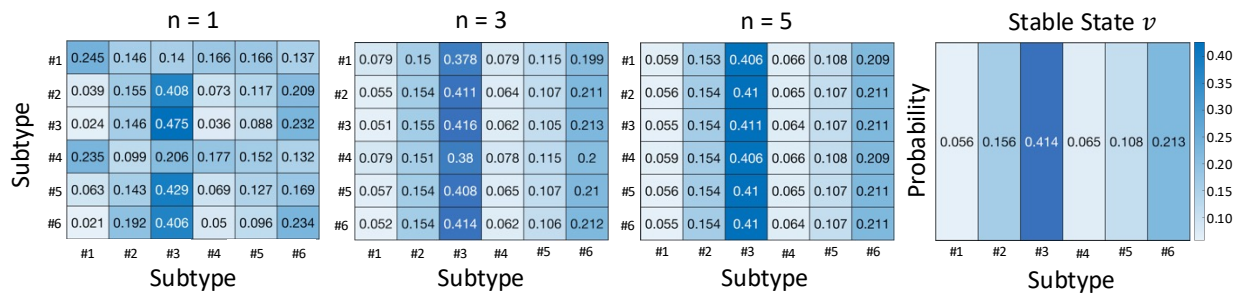

**Fig. 6. Heatmaps of subtype transition matrix at different time steps ($n$).** The number and shade in entry $i^{th}$ row and $j^{th}$ column represent the probability of transiting from subtype #$i$ to subtype #$j$ at a given time.

## 5 CONCLUSIONS

In this paper, we propose a pathology steered stratification network to fill the domain knowledge gap in the current study of AD subtype identification. Our proposed method integrates (1) a state-of-art reaction-diffusion model that can identify causality and mechanisms underlying Alzheimer's and characterize the spatiotemporal dynamics of AT[N] biomarkers across brain networks; (2) a deep predictive neural network that can learn representative features of neurodegeneration trajectories and stratify the aging population into fine-grained subtypes with distinct pathological profiles. Compared to other methods like K-means, our PSSN achieves the highest inter-cluster heterogeneity and intra-cluster homogeneity. Importantly, we discovered six distinct subtypes spread across the neurodegeneration spectrum, where each subtype shows consistency between neurobiological burden and symptom profiles, indicating the potential of our PSSN to disentangle AD heterogeneity. An evolutionary disease graph is presented as a general guideline for the probable state transition of subtypes. Altogether, our method provides a systematic way of subtype identification with joint consideration of existing pathological knowledge, physiological imaging data, and psychological examinations, which can assist pre-clinical AD prognosis and may be applied to other neurodegenerative studies facing similar therapeutic challenges.

## CONFLICT OF INTEREST

The authors have no conflict of interest to report.

## ACKNOWLEDGMENTS

The work is supported by NIH R03AG073927 and Wake Forest University Pilot Grant Funding.

## DATA AVAILABILITY STATEMENT

The data that support the findings of this study are openly available in the ADNI database at https://adni.loni.usc.edu/. The source code used to produce the results and analyses presented in this manuscript is available on a GitHub repository at https://github.com/chenm19/PSSN.